\documentstyle{article}
\begin{document}
\title{Differential Form Approach for Stationary Axisymmetric Maxwell Fields in General
Relativity}
\author{L. Fern\'andez-Jambrina and F.J. Chinea\\
Departamento de F\'{\i}sica Te\'orica II,
\\Facultad de
Ciencias F\'{\i}sicas,
\\Universidad Complutense 
\\28040-Madrid, Spain}
\maketitle 
\begin{abstract}
A formulation for stationary axisymmetric electromagnetic
fields in general relativity is derived by casting them into the form of an
anisotropic fluid. Several simplifications of the formalism are
carried out in order to analyze different features of the fields, such as the
derivation of electromagnetic sources for the Maxwell field in the form of
thin layers, construction of new solutions, and generation techniques.  
\\
\noindent PACS: 04.20.Cv, 04.20.Jb, 04.40.+c
 \end{abstract}

\section{Introduction} 

The study of coupled stationary axisymmetric
electromagnetic and gravitational fields is of great interest in astrophysics
in order to provide a description for the exterior of steadily rotating
objects. Although the whole system of equations can be reduced to two complex
partial differential equations \cite{Ernst} and special techniques have been
 developed to generate solutions, for instance the ones reviewed in
\cite{Corn}, \cite{Esc} for vacuum fields, we still lack exact solutions whose
physical interpretation is appealing. 

In section 2 we shall
extend the  exterior differential system formalism presented in \cite{ch} and
\cite{tes} for perfect fluids to electromagnetic fields. The Maxwell field
will be presented as an anisotropic fluid. Null fields will not be considered
since they have a null Killing vector \cite{Gur} and have already been studied.
The main advantage of using differential forms is that we are not forced to
choose coordinates from the beginning and therefore there is more freedom for
solving the equations and simplifying the system in different ways.  In section
3 a simplification will be carried out on this Maxwell fluid to cast it into a
shear-free form. Advantage will be taken from it to construct magnetic dipole
thin layers as surface sources for the magnetic field of static (among other)
electrovacs and to extend previous results on angular momentum densities for
vacuum metrics to some electrovacs.  Section 4 is devoted to another
simplification of the formulation. This time the Maxwell fluid will take the
form of an irrotational fluid. An extension of some results obtained in
canonical coordinates is provided. In section 5 it is performed a
simplification of different nature: The Ricci tensor is diagonalized and a
family of new electrovacs is constructed. 

\section{Exterior differential
system for stationary axisymmetric electrovacs}

\subsection{Cartan's first structure equations}

We shall follow the approach developed in \cite{ch} and \cite{tes} for the
kinematical quantities and define an orthonormal vierbein
$\{\theta^0,\,\theta^1,\,\theta^2,\,\theta^3\}$ where 
$\{\theta^0,\,\theta^1\}$ lie on the space spanned by the orbits of the
isometries $\{\partial_t,\partial_\phi\}$ whereas $\{\theta^2,\,\theta^3\}$
will lie on the orthogonal space. In order to partially diagonalize the
electromagnetic stress tensor we shall choose $\theta^0$ parallel to the
direction of its timelike eigenvector as the timelike leg of the tetrad. The
four eigenvalues of the stress tensor have the same absolute value and one of
them is negative. This one corresponds generically to the eigenvector which
lies in the direction of the magnetic and electric field in a frame in which
they are parallel or, if the Lorentz invariant $E\cdot B$ is zero, then the
respective eigenvector lies in the direction of the electric (magnetic) field
in a frame in which the magnetic (electric) field vanishes. Therefore we shall
be able to view the electromagnetic field as an anisotropic fluid with the
absolute value of the pressure equal to the density and consider $u=-\theta^0$
as its velocity form. Imposing that the tetrad be torsion-free we get the
following Cartan equations:
 \begin{equation} 
d u = a \wedge u + w \wedge \theta^1 
\end{equation} 
\begin{equation}
 d \theta ^1 = (b - a) \wedge \theta ^1 + s\wedge u
\end{equation}
\begin{equation}
d\theta^2=-\nu\wedge\theta^3 
\end{equation}
\begin{equation}
d\theta^3=\nu\wedge\theta^2 
\end{equation}
 where $a$ is the acceleration of the `fluid', $w=*\omega$ 
($\omega$ is the vorticity form derived from $u$ and $*$ stands for the
two-dimensional Hodge dual in the space orthogonal to the orbits of the Killing
vectors $\{\partial_t,\partial_\phi\}$), $s$ is a one-form associated with the
shear of the fluid, $b=d\ln\rho$ ($\rho$ is the radial pseudocylindrical Weyl
coordinate) and $\nu$ is just a connection in the two-space spanned by
$\theta^2$ and $\theta^3$. We still have a great amount of freedom for further
simplifications.

\subsection{Bianchi identities}

Bianchi identities are easily obtained by exterior differentiation of the
previous set of equations:

\begin{equation} 
d b  =  0 \label{eq:cWeyl}
\end{equation} 
\begin{equation} 
d a = w \wedge s \label{eq:acc}
\end{equation}
\begin{equation} 
d w = - (b - 2 a) \wedge w\label{eq:cvor} 
\end{equation} 
\begin{equation} 
d s = (b - 2 a) \wedge s \label{eq:csh}
\end{equation}

\subsection{Maxwell's equations}

Now we shall describe the Faraday and Maxwell electromagnetic strength
two-forms ($F$ and its four-dimensional dual $^4*F$\ ) in terms of the 
electric
and magnetic one-forms, respectively $E$ and $B$) .\ In what follows we
shall restrict ourselves to electromagnetic forms $E$ and $B$ with no
projection on the orbits of the Killing vectors:
 \begin{equation}
F=-E\wedge u+*B\wedge\theta^1
\end{equation}
\begin{equation}
^4*F=B\wedge u +*E\wedge\theta^1
\end{equation}

Therefore, the Maxwell vacuum equations $dF=0$ and $d\ ^4*F=0$ imply:
\begin{equation}
dE-E\wedge a+*B\wedge s=0\label{eq:irrot}
\end{equation}
\begin{equation}
d*B+E\wedge w-*B\wedge(b-a)=0\label{eq:sol}
\end{equation}
\begin{equation}
d*E-B\wedge w-*E\wedge(b-a)=0\label{eq:Gauss}
\end{equation}
\begin{equation}
dB-B\wedge a-*E\wedge s=0\label{eq:Ampere}
\end{equation}
These equations are not a consequence of the integrability conditions for the
Einstein equations unless the Lorentz scalar $E\cdot B$ is different from
zero. These equations can be written in a more compact expression defining a
complex one-form $f$:

\begin{equation}
f=E+iB
\end{equation}
\begin{equation}
df=-a\wedge f-is\wedge *f\label{eq:mrot}
\end{equation}
\begin{equation}
d*f=(a-b)\wedge *f+iw\wedge f\label{eq:mdiv}
\end{equation}

\subsection{Einstein's equations}

Considering that the electromagnetic stress tensor has the following
expression: 
\begin{equation}
{T^a}_b=\frac{1}{4\pi}\{F^{ac}F_{bc}-\frac{1}{4}\delta^a_bF^{cd}F_{cd}\},
\end {equation}
the Einstein equations take the general form:

\begin{equation}
d * (w -s) + 2 a \wedge * w
+ 2 (a - b) \wedge * s =4E\wedge B=2if\wedge\bar{f}\label{eq:vor}
\end{equation} 
\begin{equation}
d * a + b \wedge * a + \frac{1}{2} w \wedge * w
- \frac{1}{2} s \wedge * s = E\wedge *E+B\wedge *B=\bar f\wedge
*f\label{eq:Ray} 
\end{equation}
\begin{equation}
d * b + b \wedge * b = 0\label{eq:Weyl}
\end{equation}
\begin{eqnarray}
d \tilde b +b\wedge\tilde b- \frac{1}{2} (s - w) \wedge (\tilde{s} -
\tilde{w})+ 2a \wedge \tilde{a}- 2b \wedge \tilde{a}  
+\nonumber\\+ 2\nu\wedge *\tilde b=2E\wedge \tilde E+2B\wedge \tilde
B=2\bar{f}\wedge\tilde{f} \label{eq:sob}\end{eqnarray}
\begin{eqnarray}
d  *\tilde b + b \wedge *\tilde b - \frac{1}{2} (s - w) \wedge * (\tilde{s}
 - \tilde{w})
+ 2a \wedge * \tilde{a}- 2b \wedge * \tilde{a} 
-\nonumber\\-2\nu\wedge\tilde b=2E\wedge *\tilde E+2B\wedge *\tilde B=2\bar
f\wedge *\tilde {f}\label{eq:sob2}
 \end{eqnarray}
\begin{equation}
d\nu+a\wedge *b-a\wedge *a+\frac{1}{4} (s - w) \wedge * (s-w)=0\label{eq:dif}
\end{equation}

The symbol $\sim$ denotes the linear transformation in the
$\{\theta^2,\theta^3\}$\,-space defined by the following expression:
\begin{eqnarray}
\tilde{\theta}^2=\theta^2\ \ \ \ \ \ \ \ \ \ \tilde{\theta}^3=-\theta^3
\end{eqnarray} 

As in the perfect fluid case,  equations \ref{eq:sob} to \ref{eq:dif} can be
left for the end, since the connection $\nu$ can be algebraically obtained from
them and then equation \ref{eq:dif} is automatically satisfied.

\section {The electromagnetic field as a rigidly rotating fluid}

\subsection{Ernst's equations}

As we have already said, our exterior system can be simplified in many 
ways. This freedom comes from the possibility of describing the
electromagnetic field in different reference frames preserving the structure
of the equations. We can shift from one frame to another performing a local
boost or rotation or a combination of both. 

One boost that can always be applied is the one which takes us from our
original frame to another in which the velocity $u$ is shear-free ($s=0$). This
transformation is described by the following equations:

\begin{equation}
u'=\cosh\lambda\ u+\sinh\lambda\ \theta^1
\end{equation}
\begin{equation}
{\theta^1}'=\sinh\lambda\ u+\cosh\lambda\ \theta^1
\end{equation}
\begin{equation}
d\lambda=\cosh\lambda\ \sinh\lambda\ (b'-2a')+\cosh^2\lambda\
s'-\sinh^2\lambda\  w'
\end{equation}

The electromagnetic field changes in the following way:

\begin{equation} 
E=\cosh\lambda\,E'-\sinh\lambda\,*B'
\end{equation}

\begin{equation} 
B=\cosh\lambda\,B'+ \sinh\lambda\,*E'
\end{equation}

This transformation is compatible with the whole set of equations as it
happened in the vacuum case \cite{ch}. Therefore there is no loss of
generality in taking $s=0$ from the beginning.

We can formally integrate equations \ref{eq:cWeyl} to \ref{eq:cvor} and
\ref{eq:mrot} to get:

\begin{equation}
a=dU\ \ \ \ \ \ \ b=d\ln\rho \label{eq:int}
\end{equation}

\begin{equation}
w=\rho^{-1}e^{2U}dA\label{eq:w}
\end{equation}

\begin{equation}
f=-e^{-U}d\Phi\label{eq:f}
\end{equation}

From equation \ref{eq:vor} we obtain the exact differential that
defines the twist potential $\chi$ \cite{Ernst}:

\begin{equation}
d\chi=e^{2U}*w+2i\bar\Phi d\Phi\label{eq:twist}
\end{equation}
and construct the Ernst potential $\varepsilon$ in the following way:

\begin{equation}
d\varepsilon=de^{2U}+id\chi\ \ \ \ \ \ \
{\cal R}\varepsilon=e^{2U}-\bar\Phi\Phi\label{eq:epsilon} 
\end{equation}

Substituting \ref{eq:f} in \ref{eq:mdiv} we get an elliptic
equation for the scalar complex electromagnetic potential $\Phi$:

\begin{equation}
d*d\Phi+b\wedge *d\Phi=\frac{1}{{\cal
R}\varepsilon+\Phi\bar\Phi}\{d\varepsilon+2\bar\Phi d\Phi\}\wedge
*d\Phi\label{eq:Laplace}
 \end{equation}

And if we combine  the previous equation with the Raychaudhuri equation
(\ref{eq:Ray}), and \ref{eq:twist} with \ref{eq:int} and
\ref{eq:cvor}, we obtain another second order elliptic equation, namely the
Ernst equation:

\begin{equation}
d*d\varepsilon+b\wedge *d\varepsilon=\frac{1}{{\cal
R}\varepsilon+\Phi\bar\Phi}\{d\varepsilon+2\bar\Phi d\Phi\}\wedge
*d\varepsilon\label{eq:Ernst}
 \end{equation}

The whole set of equations for stationary axisymmetric electrovacs reduces to
these two complex equations for the potentials $\varepsilon$ and $\Phi$
\cite{Ernst}.

The equations \ref{eq:cWeyl} and \ref{eq:Weyl} just introduce two functions
$\rho$ and $z$, the Weyl pseudocylindrical coordinates, that are related by
the Hodge dual of their differentials:

\begin{equation}
*d\rho=-dz
\end{equation}

And therefore the metric takes the following form, after integrating the Cartan
equations:

\begin{equation}ds^{2}=-e^{2U}(dt-Ad\phi)^{2}+e^{-2U}[e^{2k}(d\rho^{2}+dz^
{2})+\rho^2d\phi^{2}]\label{eq:can}
\end{equation}

\subsection{Construction of electromagnetic dipole surface densities}

This formalism can be used to obtain electric and magnetic dipole
surface densities for the sources of some  asymptotically flat stationary
axially symmetric solutions of the Einstein-Maxwell equations in a similar way
as it was done in \cite{first} to construct angular momentum densities for
vacuum metrics. The formulae derived by Israel in \cite{is1}
cannot be applied since they only yield non-vanishing expressions for the mass,
charge and angular momentum, but not for the magnetic momentum.

We shall only consider static fields and also nonstatic metrics for
which $w\wedge f=0$ -if we want to calculate only magnetic (electric) moment
densities, we shall only need $w\wedge E=0$ \ ($w\wedge B=0$)-. By 
asymptotically flat we mean that the metric has the following
behavior at infinity ($r\rightarrow\infty$) in some coordinates
$(t,r,\theta,\phi)$:

\begin{eqnarray}ds^2=-(1-\frac{2m}{r})(dt+\frac{2J\sin^{2}\theta}{r}d\phi)^2+
\nonumber\\+(1+\frac{2m}{r})[dr^2+r^2(d\theta^2+
\sin^2\theta d\phi^2)] \end{eqnarray}
where $m$ is the source's total mass and $J$ is the total angular
momentum. Therefore the metric functions $e^{2U}$ and $A$ and the twist
potential $\chi$ (since $A$ and $\chi$ are related by the dual of the
vorticity one-form) take this form at infinity:

\begin{equation}
e^{2U}=1-\frac{2m}{r}+O\left(\frac{1}{r^2}\right)\ \ \ \ \ \ \ A=-\frac{2J
\sin^{2}\theta}{r}+O\left(\frac{1}{r^2}\right)
\end{equation}

\begin{equation}
\chi=-\frac{2J\cos\theta}{r^2}+O\left(\frac{1}{r^3}\right)
\end{equation}

To calculate the magnetic dipole we shall require that the
magnetic scalar potential, that is the imaginary part $V$ of
the potential $\Phi$ must be asymptotically dipolar:

\begin{equation}
V=\frac{M\cos\theta}{r^2}+ O\left(\frac{1}{r^3}\right)
\end{equation}

As it happens in flat spacetime, the magnetic scalar potential can be
defined only outside the sources. If we have a sheet of dipoles as a source,
the limit values of the potential on either side of the surface will be
different and this discontinuity reveals the existence of the source (cfr. for
instance \cite{Kel}). Our purpose will be to extend this result to curved
spacetimes.

Following the approach developed in \cite{first}, we integrate equation
\ref{eq:sol} with the previously stated restrictions, that combined with
\ref{eq:f} provides us with two expressions for the magnetic field:

\begin {equation}
B={\cal I} \{f\}=-e^{-U}dV=-\rho^{-1}e^U*dW
\end{equation} 

\begin{equation}
V={\cal I}\{\Phi\}
\end{equation}

This function $W$ has the following asymptotic expression due to the 
behavior of $V$ at infinity:

\begin{equation}
W=\frac{M\sin^2\theta}{r}+O\left(\frac{1}{r^2}\right)
\end{equation}

And now we shall integrate the scalar product of the magnetic field $B$ and
the differential of a function $Z(r,\theta)$ to be determined. The domain of
integration will be the space $V_3$ orthogonal to the congruence defined by
$u$. The metric on this space is the projection $g=^4g+u\otimes u$. Using both
expressions for $B$ we get:

\begin{eqnarray}
0=\int_{V_3}\sqrt{g}<[B+*(*B)],dZ>dx^1dx^2dx^3=
\nonumber\\  =\int_{V_3}\sqrt{g}\{-e^{-U}g^{\mu\nu}\partial_{\mu}V
\partial_{\nu}Z+\rho^{-1}e^U\varepsilon^{\mu\nu}\partial_{\mu}W
\partial_{\nu}Z\}dx^1dx^2dx^3\label{eq:geo}
\end{eqnarray} 

In order to express the integrand as a total derivative we choose $Z$ so that
it fulfills the following differential equation:

\begin{equation}
\partial_{\mu}(\sqrt{g}e^{-U}g^{\mu\nu}\partial_{\nu}Z)=0\label{eq:2Lap}
\end{equation}

As an asymptotic boundary condition we impose that $Z$ behaves like
$r\cos\theta$ at infinity.

The potential $V$ is discontinuous across a closed surface $S$ if we are to
have a magnetic moment surface density, therefore we have to split the
space into two pieces $V^+_3$ and $V^-_3$, respectively the outer and inner
part of $V_3$. If the surface $S$ is open, it can be extended to a
closed one, taking the value zero for the discontinuity on the closure.
$V^+$ and $V^-$ shall denote the limit values of $V$ on either side
of the surface.

Since the Levi-Civit\`a tensor on the space orthogonal to the orbits of the
Killing fields is $\varepsilon^{\mu\nu}=e^{2(U-k)}[\mu\nu]$ the integrand in
\ref{eq:geo} can be written as:

\begin{equation}
\partial_{\mu}\{-\sqrt{g}e^{-U}g^{\mu\nu}V\partial_{\nu}Z+W[\mu\nu]
\partial_{\nu}Z\}
\end{equation}

We have the desired total derivative and we can express the integral as a
surface integral on the boundary $\Sigma=\partial V^+_3\cup\partial V^-_3$
with unitary normal $n$. The boundary of
$V_{3}^+$ consists of $S$ and the sphere at infinity $S^2(\infty)$ and the
boundary of $V_{3}^-$ is just $S$. Taking into account the required behavior
of the metric functions and the magnetic potential at infinity, the integral
on $S^2(\infty)$ yields:

\begin{eqnarray}
\int_SdS[V]e^{-U}g^{\mu\nu}n_\mu\partial_{\nu}Z=4\pi M
\end{eqnarray}

And therefore we obtain the following formula relating the jump of the
magnetic potential $[V]$ across $S$ with the total magnetic moment $M$:

\begin{equation}
4\pi M=\int_SdS\ \sigma
\end{equation}

\begin{equation}
\sigma=\frac{1}{4\pi}[V]e^{-U}g^{\mu\nu}n_{\mu}\partial_{\nu}Z\label{magden}
\end{equation} 

 We can consequently infer that $\sigma$ is the source's magnetic moment density
for the Maxwell field. 

Similar calculations can be done with the real part of the
electromagnetic potential $\Phi$ to compute electric dipole surface densities.

\subsection{Angular momentum surface densities}

The formalism developed in \cite{first} for interpreting discontinuities of the
twist potential of asymptotically flat stationary axially symmetric vacuum
metrics can be extended in a straightforward way to cope with electrovac
solutions with $E\wedge B=0$, since the only equations
involved are the Bianchi equations and the equation for the
vorticity, \ref{eq:vor}, and these remain the same as in the vacuum case if we
impose the mentioned restriction. Hence the angular momentum surface density
for stationary axisymmetric vacuum metrics and electrovacs with parallel
magnetic and electric fields has the following expression in terms of the jump
$[\chi]$ of the twist potential:

\begin{equation}
\sigma_{rot}=-\frac{1}{8\pi}[\chi]e^{-3U}g^{\mu\nu}n_\mu\partial_{\nu}Z_{rot}
\end{equation}
where the function $Z_{rot}$ is asymptotically $r\cos\theta$ and satisfies the
following differential equation:

\begin{equation}
\partial_{\mu}(\sqrt{g}e^{-3U}g^{\mu\nu}\partial_{\nu}Z_{rot})=0\label{eq:lap}
\end{equation} 

\subsection{Bonnor's transformation}

One of the peculiarities of this formalism for constructing surface densities is
its behavior under Bonnor transformations. These transformations
are  a method for generating magnetostatic electrovacs from stationary
nonstatic metrics \cite{Bon1}. As the solutions of the Einstein-Maxwell
equations are fully characterized by the Ernst and electromagnetic potentials,
it will suffice to give the rules that determine the potentials for the
electrovac:

\begin{equation}
\varepsilon=\varepsilon_{vac}\,\bar{\varepsilon}_{vac}
\end{equation}

\begin{equation}
V=i\chi_{vac}
\end{equation}

In order to have a real solution the magnetic potential has to be real,
therefore the parameters have to be rearranged with a complex
transformation to achieve this purpose. 

From these expressions it is clear that
$e^U=e^{2U}_{vac}$. Since $Z$ is
 a function only of $\rho$ and $z$, in canonical coordinates, the only
components of the three-metric $g$ of $V_3$ that appear in the
differential equation \ref{eq:2Lap} are $g^{\rho\rho}=g^{zz}=e^{2U-2k}$. The
square root of the determinant of this metric is $\sqrt{g}=e^{-3U+2k}\rho$  and
therefore equation  \ref{eq:2Lap} takes this form in terms of the original
vacuum metric functions after rearranging the parameters:

\begin{eqnarray}
\partial_{\mu}(\sqrt{g}e^{-U}g^{\mu\nu}\partial_{\nu}Z)&=&\partial_{\mu}(e^{-2U}
\rho\delta_{\mu\nu}\partial_{\nu}Z)=\nonumber\\
\partial_{\mu}(e^{-4U}_{vac}\,
\rho\delta_{\mu\nu}\partial_{\nu}Z)&=&\partial_{\mu}(\sqrt{g_{vac}}\,
(e^{-3U})_{vac}\,
g^{\mu\nu}_{vac}\,\partial_{\nu}Z)=0
 \end{eqnarray}

That is, the function $Z$ needed for the construction of the magnetic dipole
surface density satisfies the same differential equation as the $Z_{rot}$
involved in the calculation of the angular momentum density for the original
vacuum metric $g_{vac}$. Therefore the same function is valid for both
solutions, once we have reinterpreted the parameters, and the formalism is
compatible with the Bonnor transformation. 

\subsection{An example: Bonnor's massive magnetic dipole}

As an example of how this formalism works, we shall calculate the magnetic
source for Bonnor's massive magnetic dipole \cite{Bon2}. This is a Bonnor
transformation of the Kerr metric and therefore we can use the $Z$ function
obtained in \cite{first} for its source's angular momentum density.

Bonnor's metric has the following form:

\begin{eqnarray}
ds^2=-(1-\frac{2mr}{r^2-a^2\cos^2\theta})^2dt^2+\nonumber\\+
(1-\frac{2mr}{r^2-a^2\cos^2\theta})^{-2}\left\{
 (r^2-a^2-2mr)\sin^2\theta d\phi^2+\right.\nonumber\\
\left.
\frac{(r^2-a^2\cos^2\theta-2mr)^4}{[(r-m)^2-(a^2+m^2)\cos^2\theta]^3}
(d\theta^2+\frac{dr^2}{r^2-2mr-a^2})\right\}
\end{eqnarray}
 from which we can read the necessary
metric functions:

\begin{equation}
e^{2U}=\left(1-\frac{2mr}{r^2-a^2\cos^2\theta}\right)^2=1-\frac{4m}{r}+
0\left(\frac{1}{r^2}\right)
\end{equation}

The magnetic field is determined by the scalar potential:
\begin{equation}
V=\frac{2am\cos\theta}{r^2-a^2\cos^2\theta}=\frac{2am\cos\theta}{r^2}
+0\left(\frac{1}{r^4}\right)
\end{equation}

Therefore both the gravitational and magnetic fields fulfill the required
asymptotic conditions with mass equal to $2m$ and magnetic moment $2am$.

If we consider the metric to be written in a sort of oblate spheroidal
coordinates as it is done in \cite{is1} for  the Kerr metric, then 
events on the surface $r=0$ with polar angle $\theta$ have to be identified
with events with polar angle equal to $\pi-\theta$. To avoid
double-counting these points, the $\theta$ coordinate will range from $0$ to
$\pi/2$ on this surface. Therefore the function $\cos\theta$ suffers a
discontinuity upon crossing $r=0$, since it shifts sign from positive in the
upper subspace ($r>0$, $0\leq\theta<\pi/2$) to negative in the lower subspace (
$r>0$, $\pi/2\leq\theta<\pi$).

Hence the magnetic potential $V={\cal I}\{\Phi\}$ is discontinuous on the
surface $r=0$. The difference between the values taken on the upper and the
lower side of the surface is:

\begin{equation}
[V]=-\frac{4m}{a\cos\theta}
\end{equation}

The $Z$ function satisfying \ref{eq:2Lap} can be obtained from the $Z_{rot}$ for
Kerr, as it was shown in the previous section, taking into account that
$a_{Kerr}=ia$. Hence,

\begin{equation}
Z=(r-3m)\cos\theta-\frac{2a^2m(5\cos^3\theta-3\cos\theta)}
{5(r^2-a^2\cos^2\theta)}
\end{equation}

The surface element for $r=0$ is:

\begin{equation}
 dS=\frac{a^5\sin\theta\cos^4\theta}{|a^2\cos^2\theta-m^2\sin^2\theta|^{3/2}}d
\theta d\phi \end{equation}

We are ready now to write the magnetic moment surface density for the source
of Bonnor's magnetic dipole using the formula \ref{magden}:

\begin{equation}
\sigma=\frac{m}{\pi a^4}\ \frac{|a^2\cos^2\theta-m^2\sin^2\theta|^{3/2}
}{\cos^4\theta}
\end{equation}

The integral defining the total magnetic moment yields the expected result:

\begin{equation}
M=\int_S\ \sigma\ dS=\int^{2\pi}_{0}\int^{\pi/2}_{0}d\theta\ (ma\sin\theta)=2ma
\end{equation}

But unfortunately the source lies in a region where the signature of the
metric is not the usual one (for instance, the angle $\phi$ is no longer a
spacelike coordinate). Any other surface source would have to include the
region $S$ in order to cope with the discontinuities of the magnetic
potential. Therefore its interpretation as a physical material source is
dubious and one would have to resort to three-dimensional sources to hide the
unphysical regions of the electrovac metric.

\section {The electromagnetic field as an irrotational fluid}

\subsection{Ernst's equations}

Instead of performing a boost in order to cancel the shear one-form $s$, as it
was done in the previous section, we can always write our equations in a frame
where the dual of the vorticity $w$ is zero. This is also compatible with our
system of equations, as it happened for vacuum fields \cite{ch}, \cite{tes}. Of
course, this would not be possible in general for a perfect fluid. The
differential equation defining the $\lambda$ parameter of the boost is now:

\begin{equation}
d\lambda=\cosh\lambda\ \sinh\lambda\ (2a'-b')+\cosh^2\lambda\
w'-\sinh^2\lambda\  s'
\end{equation}

This amounts to many simplifications, as it happened in the shear-free
formalism. The equations \ref{eq:cWeyl} to \ref{eq:csh} and
\ref{eq:mdiv} can be integrated to get:

\begin{equation}
a=dU\ \ \ \ \ \ \ b=d\ln\rho \label{eq:iint}
\end{equation}

\begin{equation}
s=\rho e^{-2U}dC\label{eq:s}
\end{equation}

\begin{equation}
*f=-e^{U}\rho^{-1}d\Psi\label{eq:df}
\end{equation}

From equation \ref{eq:vor} we get a exact differential that
can be used to define what could be called 'deformation potential' $\psi$ in
analogy with the twist potential:

\begin{equation}
d\psi=\rho^{2} e^{-2U}*s-2i\bar\Psi d\Psi\label{eq:deform}
\end{equation}
and construct another complex potential $\eta$ similar to the Ernst potential:

\begin{equation}
d\eta=d(-\rho^2e^{-2U})-id\psi\ \ \ \ \ \ \
{\cal R}\{\eta\}=-\rho^2e^{-2U}-\bar\Psi\Psi\label{eq:eta} 
\end{equation}

If we substitute \ref{eq:df} in \ref{eq:mrot} another generalization of
the Laplace equation is achieved, this time for the potential $\Psi$:

\begin{equation}
d*d\Psi+b\wedge *d\Psi=\frac{1}{{\cal
R}\eta+\Phi\bar\Phi}\{d\eta+2\bar\Psi d\Psi\}\wedge
*d\Psi\label{eq:dLaplace}
 \end{equation}

The analog of the Ernst equation can be obtained in this irrotational
formalism in a similar way taking into account the latter equation, the
Raychaudhuri equation (\ref{eq:Ray}) and the equations \ref{eq:deform},
\ref{eq:iint} and \ref{eq:csh}:

\begin{equation}
d*d\eta+b\wedge *d\eta=\frac{1}{{\cal
R}\eta+\Psi\bar\Psi}\{d\eta+2\bar\Psi d\Psi\}\wedge
*d\eta\label{eq:dErnst}
 \end{equation}

These two equations, together with the coordinate condition $*d\rho=dz$,
comprise the whole problem of calculating stationary axially symmetric
electrovac solutions. In this formalism the metric takes a form different from
the canonical one:

\begin{equation}ds^{2}=-e^{2U}dt^{2}+e^{-2U}[e^{2k}(d\rho^{2}+dz^
{2})+\rho^2(d\phi+Cdt)^2]\label{eq:dcan}
\end{equation}

Both sets of equations, the Ernst equations (\ref{eq:Laplace}
and \ref{eq:Ernst}) and these new Ernst-like equations, look the same
exchanging $\varepsilon$ for $\eta$ and $\Phi$ for $\Psi$ as it happened in
the vacuum case \cite{ch}, \cite{tes}. However, there is a
difference that did not occur for vacui and that is hidden behind the minus
sign that it is written before the norm of the rotation Killing vector,
$\|\partial_\phi\|=\rho^2e^{-2U}$, as it appears in the definition for the
$\eta$ potential. This sign enforces the real part of $\eta$ to be negative if
it has to deal with a physical situation. Otherwise the rotation Killing
vector would be timelike. On the other hand $\varepsilon$ can be either
positive or negative. Besides, the conditions on the potentials for the metric
to be asymptotically flat are rather different:

\begin{equation}
\varepsilon=1+0\left(\frac{1}{r}\right)\label{asEr}
\end{equation}

\begin{equation}
\eta=-r^2\sin^2\theta+0(r)
\end{equation}

Although the equations have been cast in the same form, in fact the potentials
have very different meaning.

One could think of the following transformation in order to generate new
stationary axially symmetric electrovacs as the electromagnetic generalization
of the transformation introduced in \cite{ch} and \cite{tes}:

\begin{equation}
\varepsilon\longrightarrow\eta
\end{equation}

\begin{equation}
\Phi\longrightarrow\Psi
\end{equation}

But this would lead to unphysical metrics unless the seed metric has $(+---)$
signature, since this transformation induces the following exchange between
the norms of the Killing vectors:

\begin{equation}
e^{2U}\longrightarrow-\rho^2e^{-2U}
\end{equation}

It could be used then to generate new solutions from the forbidden regions of
other spacetimes.

\subsection{Two families of `irrotational' solutions}

The first attempt for constructing new solutions of the Einstein-Maxwell
equations with the irrotational formalism could be to make use of the
ans\"atze that have been profitable for solving the standard Ernst equation.
For a start it can be shown that the Weyl family of  static electrovacs
\cite{Weyl} can be extended to yield new solutions.

The Weyl ansatz, in terms of the Ernst potentials, assumes a functional
dependence between $\varepsilon$ and $\Phi$, that now are real (Since the
metric is static, the twist potential can be taken to be null): 

\begin{equation}
\varepsilon=\varepsilon(\Phi)
\end{equation}

This ansatz leads to two equations, that can be integrated, considering the
asymptotic flatness condition for the Ernst potential (\ref{asEr}):

\begin{equation}
\ddot\varepsilon(\Phi)=0\Rightarrow e^{2U}=1-2c\Phi+\Phi^2
\end{equation}
where $c$ is a constant.

\begin{equation}
d*d\Phi+b\wedge*d\Phi=\frac{-2c+2\Phi}{1-2c\Phi+\Phi^2}d\Phi\wedge*d\Phi
\end{equation}
that can be solved in terms of a function $Y$, solution of the reduced Laplace
equation:

\begin{equation}
d*dY+b\wedge*dY=0
\end{equation}
via the following relation:

\begin{equation}
\Phi=\left\{
\begin{array}{l}
-\sqrt{c^2-1}\coth Y+c\\
-Y^{-1}+c\\
-\sqrt{1-c^2}\cot Y+c
\end{array}
\right.e^{2U}=\left\{
\begin{array}{lr}
(c^2-1)\sinh^{-2}Y&c^2>1\\
Y^{-2}&c^2=1\\
(1-c^2)\sin^{-2}Y&c^2<1
\end{array}
\right.
\end{equation}

A similar ansatz can be applied in the irrotational formulation of static
axisymmetric electrovacs:

\begin{equation}
\eta=\eta(\Psi)
\end{equation}

The equations that this ansatz yields are entirely similar to those of the
Weyl ansatz:

\begin{equation}
\ddot\eta(\Psi)=0\Rightarrow \rho^2e^{-2U}=c'+2c\Psi-\Psi^2
\end{equation}

\begin{equation}
d*d\Psi+b\wedge*d\Psi=\frac{2c-2\Psi}{c'+2c\Psi-\Psi^2}d\Psi\wedge*d\Psi
\label{Weyldual}
\end{equation}

For physical reasons we want $c'+2c\Psi-\Psi^2$ to be positive and this can
only be achieved if $c'+c^2>0$. With this restriction in mind the physical
cases to be considered reduce to just one:

\begin{equation}
\Psi=c+\sqrt{c'+c^2}\tanh Y\ \ \ \ \ \ \ \ \rho^2e^{-2U}=(c'+c^2)\cosh^{-2}Y
\end{equation}
where $Y$ is again a solution of the reduced Laplace equation.

From the previous equation we learn that $ \rho^2e^{-2U}$ is upperly bounded
by $c'+c^2$ and hence these solutions cannot be asymptotically flat at
infinity, since the norm of the rotational Killing vector is expected to
diverge at infinity. This family could be used to represent an electrovac
spacetime surrounded by a material source to which it should be matched.

Another ansatz that can be extended to this irrotational formalism is that of
Papapetrou \cite{Papa} for vacuum metrics. Following \cite{Kramer} we can
write this ansatz as an assumption of functional
dependence between the real and the imaginary parts of the Ernst potential:

\begin {equation}
\varepsilon= e^{2U}+i\chi\ \ \ \ \ \ \ \ U=U(\chi)
\end{equation}

The real and imaginary parts of the Ernst equation imply:

\begin{equation}
e^{4U}=c'+2c\chi-\chi^2
\end{equation}
where $c$ and $c'$ are constants.

\begin{equation}
d*d\chi+b\wedge*d\chi=\frac{2c-2\chi}{c'+2c\chi-\chi^2}d\chi\wedge*d\chi
\end{equation}
which is similar to the one we have integrated for the extension of the Weyl
solution (\ref{Weyldual}). Therefore the twist potential has the following
expression in terms of an arbitrary solution of the reduced Laplace equation: 

\begin{equation}
\chi=c+\sqrt{c'+c^2}\tanh Y
\end{equation}

The restriction $c'+c^2>0$ has to be imposed again for physical reasons. The
Ernst potential takes the form:

\begin{equation}
\varepsilon=\sqrt{c'+c^2}(\cosh^{-1} Y +i\tanh Y)+ic
\end{equation}

If we apply this ansatz to the irrotational formalism,
$\rho^2e^{-2U}=g(\psi)$, instead of to the habitual shear-free formalism, then
the equations have the same form and therefore we have a different family of
solutions:

\begin{equation}
\eta=\sqrt{c'+c^2}(\cosh^{-1} Y +i\tanh Y)+ic
\end{equation}

The same considerations for the asymptotic behavior of the extension of the
Weyl family apply to this extension of the Papapetrou class.

\section{Diagonalization of the stress tensor}

Classical nonnull electromagnetics fields can be put in a frame where their
stress tensor is diagonal by performing a Lorentz transformation. It implies
taking the Poynting vector to be zero by letting the electric and magnetic field
become parallel (if the Lorentz invariant $E\cdot B$ is different from zero) or
by cancelling either the electric or the magnetic field (if $E\cdot B=0$ and
the other Lorentz invariant $\|E\|^2-\|B\|^2$ is respectively negative or
positive).

The same simplification can be applied without loss of generality in general
relativity. This amounts to taking $E\wedge B=0$ in the exterior system and
therefore the Einstein equations for the difference of the vorticity and the
shear and for the dual of the acceleration remain as follows:

\begin{equation}
d * (w -s) + 2 a \wedge * w
+ 2 (a - b) \wedge * s =0\label{eq:newvor}
\end{equation} 
\begin{equation}
d * a + b \wedge * a + \frac{1}{2} w \wedge * w
- \frac{1}{2} s \wedge * s = E\wedge *E+B\wedge *B=\bar f\wedge
*f
\end{equation}

\subsection{`Seminull' Maxwell fields}

In this section we shall call `seminull' the Maxwell fields that have one 
of the Lorentz invariants identically zero in opposition to null and generic
fields. There are three different cases:

As we have already mentioned, when $E\cdot B=0$ and $E^2-B^2<0$ the equations
can be expressed in a frame where $E=0$. We shall write only the relevant
equations where the electromagnetic field appears: The Raychaudhuri
equation (\ref{eq:Ray}) and the Maxwell equations:

\begin{equation}
d * a + b \wedge * a + \frac{1}{2} w \wedge * w
- \frac{1}{2} s \wedge * s = B\wedge *B\label{nray}
\end{equation}

\begin{equation}
dB-B\wedge a=0\label{eq:nirrot}
\end{equation}

\begin{equation}
d*B-*B\wedge(b-a)=0\label{eq:nGauss}
\end{equation}

\begin{equation}
B\wedge w=0\label{eq:nsol}
\end{equation}

\begin{equation}
*B\wedge s=0\label{eq:nAmpere}
\end{equation}

From the last two equations we get as a constraint that the vorticity and
shear one-forms have to be orthogonal.

A similar case happens when $E\cdot B=0$ and $E^2-B^2>0$. Now it is $B$ the
field that can be taken to be zero. The resulting equations are the same as
\ref{nray}-\ref{eq:nAmpere} after changing $B$ for $E$.

The last case of seminull fields consists of those that have $E\cdot B\neq
0$, but $E^2=B^2$. Hence, in the frame where both fields are parallel, we have
$E=\pm B$. Therefore $E$ and $B$ fulfill the same Maxwell equations, which take
the form of equations \ref{eq:nirrot} to \ref{eq:nAmpere}. The Raychaudhuri
equation becomes slightly different:

\begin{equation}
d * a + b \wedge * a + \frac{1}{2} w \wedge * w
- \frac{1}{2} s \wedge * s = 2E\wedge *E
\end{equation}

From the whole set of equations for the seminull Maxwell fields we get as a
consequence that a solution for one of the cases provides also a solution for
the other two by the following transformation rule:

\begin{equation}
B_{E^2<B^2}\longrightarrow E_{E^2>B^2}\longrightarrow\frac{1}{2}E_{E\cdot
B=0}=\pm\frac{1}{2}B_{E\cdot B=0}
\end{equation}

Although the three cases are physically different, their mathematical
description is the same.

\subsection{A family of seminull electrovacs}

In order to derive new solutions of the Einstein-Maxwell equations we shall
make use of an ansatz for the `magnetic seminull' field:

We take it to be shear-free ($s=0$) and assume the following condition on the
kinematical one-forms:

\begin{equation}
\frac{1}{2} w \wedge * w= B\wedge *B\label{con1}
\end{equation}

Due to the constraint \ref{eq:nsol}, this ansatz implies that
$B=\pm\frac{1}{\sqrt 2} w$. Therefore we have two different sets of equations
for the vorticity and its dual: 

\begin{equation} 
d w = - (b - 2 a) \wedge w\label{eq1}
\end{equation}

\begin{equation}
d * w + 2 a \wedge * w =0\label{eq2}
\end{equation} 

\begin{equation}
dw-w\wedge a=0
\end{equation}

\begin{equation}
d*w-*w\wedge(b-a)=0
\end{equation}
that will introduce a new constraint:

\begin{equation}
b=3a\label{con2}
\end{equation}

The integrability conditions for this constraint impose no new restriction since
now the equations for the acceleration form and its dual are consistent with
those governing $b$.

We already know that the equations for the one-form $b$, \ref{eq:cWeyl} and
\ref{eq:Weyl}, just introduce the Weyl coordinates $\rho$ and $z$, such that
$*d\rho=dz$. Hence, the only equations that remain to be solved are \ref{eq1}
and \ref{eq2}, that can be straightforwardly integrated:

\begin{equation}
w=\rho^{-\frac{1}{3}}dA
\end{equation}

\begin{equation}
*w=\rho^{-\frac{2}{3}}d\chi
\end{equation}
whose integrability condition in Weyl coordinates is:

\begin{equation}
A_{\rho\rho}+\frac{1}{3\rho}A_{\rho}+A_{zz}=0\label{Tricomi}
\end{equation}
which is a well-known equation in hydrodynamics, the Tricomi equation
in canonical coordinates when the parameters allow to classify it as an
elliptic equation.

The twist potential can be obtained from the solutions of the Tricomi equation
by quadratures:

\begin{equation}
\chi_\rho=\rho^\frac{1}{3}A_z
\end{equation}

\begin{equation}
\chi_z=-\rho^\frac{1}{3}A_\rho
\end{equation}

Since $a=\frac{1}{3}d\ln\rho$, the metric can be expressed as follows in
canonical coordinates:

\begin{equation}
ds^2=-c\rho^{\frac{2}{3}}(dt-Ad\phi)+c^{-1}\rho^{-\frac{2}{3}}[\rho^2d\phi^2+
e^{2k}(d\rho^2+dz^2)]
\end{equation}
$c$ being a positive constant and $k$ is obtained from the Cartan equations once
the connection $\nu$ is calculated from the equations \ref{eq:sob} and
\ref{eq:sob2}. If we write the tetrad forms $\theta^2$ and $\theta^3$ as:

\begin{equation}
\theta^2=e^Qdz\ \ \ \ \ \ \ \theta^3=e^Qd\rho
\end{equation}
then the metric function $Q=k-U$ can be integrated from this set of equations:

\begin{equation}
Q_z=-\frac{3}{2}\rho^\frac{1}{3}A_\rho A_z
\end{equation}

\begin{equation}
Q_\rho=-\frac{2}{9}\rho^{-1}-\frac{3}{4}\rho^\frac{1}{3}({A_\rho}^2- {A_z}^2)
\end{equation}
whose integrability condition is precisely the Tricomi equation \ref{Tricomi}.

From the expression for the dual of the magnetic one-form $*B$ we get the
Faraday two-form $F$:
\begin{equation}
F=\pm\frac{1}{\sqrt {2c}}d\chi\wedge d\phi
\end{equation}
and the magnetic field is :

\begin{equation}
B=\pm\frac{1}{\sqrt {2}}e^{-Q}\rho^{-\frac{1}{3}}(A_\rho\theta^2+A_z\theta^3)
\end{equation}

From the general expression for the metric it is obvious that it is not
asymptotically flat in canonical coordinates. This seems to arise from the
fact that the gravitational field obtained from the $g_{00}$ component has
cylindrical symmetry and therefore cannot be due to a compact source. The
magnetic field, however, is not cylindrical unless we assume that the metric
function $A$ does not depend on the coordinate $z$, in which case we would have
an extra Killing vector $\partial_z$.  

This family of solutions of the Einstein-Maxwell equations has got generically
only two Killing vectors and Petrov type I. It is also generically nonstatic.

\section{Discussion}

In this paper it has been introduced a new formalism for studying stationary
axially symmetric electrovacs making use of exterior differential systems of
equations and putting the electromagnetic field into the form of an
anisotropic fluid. This has been useful since we have not attached ourselves
to any special coordinate system from the beginning and therefore we have been
able to simplify our equations in different manners according to our needs.

This formulation of the coupled gravitational and electromagnetic fields has
given results both in the interpretation and in the derivation of solutions of
the Einstein-Maxwell equations. It has served to deal with sheets of dipoles as
sources for the fields and, on the other hand, to extend old families of
electrovacs and also to construct new ones. In particular, the result
concerning the magnetic dipole sources for the fields is of great interest
since they cannot be obtained by the usual formulation derived by Israel
\cite{is1}.

For a future work it would be interesting to extend further the applicability
of the techniques developed so far for interpreting the discontinuities of the
Ernst potentials so that other physical situations fit in the formalism. This
would provide a deeper insight into general relativity
and its comparison with flat-spacetime Physics.

\noindent
{\it The present work has been supported in part by DGICYT Project PB89-0142;
L.F.J. is supported by a FPI Predoctoral Scholarship from Ministerio de
Educaci\'{o}n y Ciencia (Spain). The authors wish to thank L.M.
Gonz\'alez-Romero and J.A. Ruiz-Mart\'{\i}n for valuable discussions.}

 \end{document}